# Origin and Perspectives of High Mobility in LaAlO$_3$/SrTiO$_3$ Structures

G. Herranz[1] [a], M. Basletic[2], M. Bibes[3], C. Carretero[1], E. Tafra[2], E. Jacquet[1], K. Bouzehouane[1], C. Deranlot[1], J.-L. Maurice[1], A. Hamzic[2], J.-P. Contour[1], A. Barthélémy[1], A. Fert[1]

[1]*Unité Mixte de Physique CNRS/Thales, Route départementale 128, 91767 Palaiseau Cedex, France*

[2]*Department of Physics, Faculty of Science, University of Zagreb, Bijenicka 32, P.O.B. 331, HR-10002 Zagreb, Croatia*

[3]*Institut d'Electronique Fondamentale, Université Paris-Sud, 91405 Orsay, France*

LaAlO$_3$/SrTiO$_3$ structures showing high mobility conduction have recently aroused large expectations as they might represent a major step towards the conception of all-oxide electronics devices. For the development of these technological applications a full understanding of the dimensionality and origin of the conducting electronic system is crucial. To shed light on this issue, we have investigated the magnetotransport properties of a LaAlO$_3$ layer epitaxially grown at low oxygen pressure on a TiO$_2$-terminated (001)-SrTiO$_3$ substrate. In agreement with recent reports, a low-temperature mobility of about $10^4$ cm$^2$/Vs has been found. We conclusively show that the electronic system is three-dimensional, excluding any interfacial confinement of carriers. We argue that the high-mobility conduction originates from the doping of SrTiO$_3$ with oxygen vacancies and that it extends over hundreds of microns into the SrTiO$_3$ substrate. Such high mobility SrTiO$_3$-based heterostructures have a unique potential for electronic and spintronics devices.

[a] Corresponding author: gervasi.herranz@gmail.com

The interest for the electronic properties of LaAlO$_3$/SrTiO$_3$ (LAO/STO) heterointerfaces has been recently boosted by experiments of Ohtomo et al. suggesting the existence of a high mobility conducting layer at the interface between the two insulators - a STO substrate and a LAO film grown at pressures smaller than 10$^{-4}$ mbar [1,2]. For such structures, two types of interface can be defined on the atomic level. The LAO/TiO$_2$ n-type interface was found to be conductive with low temperature Hall mobility up to $\mu_H \approx 10^4$ cm$^2$/Vs and a sheet resistance ratio R$_{sheet}$(300 K)/R$_{sheet}$(5 K) ~ 10$^3$ [1,2]. On the other hand, the p-type AlO$_2$/SrO interface is claimed to be insulating. Similar low-temperature mobilities around 10$^4$ cm$^2$/Vs have been reported for LAO/STO heterointerfaces by Siemons et al. [3]. In order to study the possible recombination of the n- and p- type interfaces, Huijben et al.[4] have grown LAO/STO/LAO and STO/LAO structures on STO substrates at 3 10$^{-5}$ mbar. They studied the transport properties of these heterostructures as a function of the thickness of the intermediate layer (either LAO or STO) and found Hall mobilities on the order of 10$^3$ cm$^2$/Vs [4].

Here we report on magnetotransport experiments performed on a LAO/STO bilayer consisting of a 20nm-thick LAO film grown by pulsed laser deposition on a 0.5 mm-thick TiO$_2$-terminated STO single-crystalline substrate. In order to determine the dimensionality of the high mobility electron gas, we have performed resistance, magnetoresistance and Hall effect measurements that enabled us to deduce the Hall mobility ($\mu_H$), sheet carrier density (n$_{sheet}$) and sheet resistance (R$_{sheet}$). We have found a metallic behavior with a mobility of about 10$^4$ cm$^2$/Vs. The period of the Shubnikov-de Haas (SdH) oscillations of the magnetoresistance is the same for parallel and perpendicular orientations of the magnetic field. This unambiguously demonstrates that the observed high-mobility transport does not occur in a nanometric interfacial slab but extends into the STO substrate over a region as thick as *hundreds of microns.* The comparison of the transport properties of LAO/STO structures with those reported for doped STO single crystals enables us to propose the existence of a universal transport behavior of systems based on n-type STO and to conclude that the high mobility transport in the LAO/STO systems is not an interface property but is due to the doping of STO by oxygen vacancies.

The magnetotransport data for our LAO/STO sample are shown in **Figs. 1** and **2**. Its mobility of $\mu_H(2K) \approx 1.8 \cdot 10^4$ cm$^2$/Vs (cf. **Fig. 2a**) and sheet resistance ratio $R_{sheet}(300K)/R_{sheet}(2K) \approx 3.8 \cdot 10^3$ (cf. **Fig. 2b** and the inset) are in good agreement with data reported in the literature for LAO/STO samples grown at low oxygen pressure [1-4] (these data will be discussed later on in more detail). Magnetoresistance curves, MR = $[R_{xx}(H)-R_{xx}(0)]/R_{xx}(0)$, were measured with the field H either perpendicular to the sample plane (PMR), or parallel to the current in the sample plane (LMR). Both magnetoresistances exhibit Shubnikov - de Haas oscillations at $T < 4$ K and magnetic fields $B \geq 6$ T (the data at 1.5 K are shown in **Fig. 1a**). After subtracting the background contribution, the SdH oscillations are even more clearly seen in **Fig. 1b**. From this figure, it is evident that the period of the oscillations does not depend on the field orientation. *Such observation of similar SdH oscillations in both configurations excludes any interfacial confinement of carriers.*

The SdH oscillations were analyzed following the protocol described in detail in Ref. 5. The Fast Fourier Transform procedure confirmed that the SdH frequency $F_{SdH}$ (which is related to the cross-sectional area $A_{ext}$ of extremal electronic orbits in *k*-space perpendicular to the applied field by $F_{SdH} = \hbar A_{ext}/2\pi e$) is the same for the PMR and LMR configurations – cf. **Fig. 1c**. This definitely demonstrates that *the electronic system is three-dimensional (3D), and that it extends homogeneously throughout the STO*. A system with a non-uniform carrier density (varying as a function of the distance from the film/substrate interface) would lead, in the PMR configuration, to a superposition of different frequencies and to a broadening or blurring of the spectrum.

Furthermore, our data enable us to estimate the thickness of the high mobility gas. The simplest approximation is to suppose a spherical Fermi surface; in this case, from $F_{SdH} \approx 11.1$ T (cf. **Fig. 1c**), we obtain $k_F \approx 1.8 \cdot 10^6$ cm$^{-1}$. The thickness of the 3D electronic system would then be much larger than the Fermi wavelength $\lambda_F = 2\pi/k_F \approx 35$ nm. To determine it more precisely, we have used our Hall resistance $R_{xy}(B)$ data (**Fig.1d**). Defining the sheet carrier density as the product of the carrier density $n$ and the thickness of the high mobility region $t_{hm}$, we get $n \cdot t_{hm} = B/(e \cdot R_{xy}) \approx 1.6 \cdot 10^{16}$ cm$^{-2}$. The carrier density $n$ can be determined *independently* by using the value of $k_F$ (derived from $F_{SdH}$); with $n = k_F^3/3\pi^2 \approx 2 \cdot 10^{17}$ cm$^{-3}$, one finally obtains $t_{hm} \approx 800$ µm. Now, if instead of an ideal spherical Fermi surface we use the more realistic k-space geometry of doped STO (as described in the Appendix of Ref. 5), we find $k_F \approx 0.9 \cdot 10^6$ cm$^{-1}$, $\lambda_F \approx 70$ nm, $n \approx 3 \cdot 10^{17}$ cm$^{-3}$ and finally $t_{hm} \approx 530$ µm. *The first conclusion is therefore that our data and their analysis unambiguously prove that the transport properties of our LAO/STO sample are due to a conducting region homogeneously extending over hundreds of µm inside the STO substrate; the conducting and high mobility behavior is then a bulk property.*

In **Fig. 2** we have collected the temperature dependence of the electronic mobility (**Fig. 2a**) and normalized electrical resistance (**Fig. 2b**) for STO bulk single crystals (doped or treated in reducing atmospheres) [6], LAO films grown on STO substrates at pressures $P_{O2} \leq 10^{-4}$ mbar (this work and Refs. 1-4), Co-doped (La,Sr)TiO$_3$ (Co-LSTO) films grown at low pressure on STO substrates (Ref. 5 – we note that we have already demonstrated that the transport properties of these samples are dominated by those of the STO substrate) and SrTiO$_{3-\delta}$ homoepitaxial films (STO$_{3-\delta}$/STO) [7]. Let us point out here that one can extract the mobility $\mu_H$ from the Hall experiments without knowing the thickness of the metallic system. This allows a reliable comparison of data from different sources. The main characteristic of both temperature dependences (mobility - **Fig. 2a** and normalized resistance - **Fig. 2b**) is the remarkable resemblance of the behavior for the LAO/STO samples and bulk STO specimens. This strongly suggests *our second conclusion that the transport properties in these two systems have the same physical origin.*

The dependence of the mobility $\mu_H$ on the carrier density for different doped STO single crystals[8, 9] is presented in **Fig. 3** (red symbols). Such a universal variation means that the carrier density is much more important than the type of dopants. The highest mobilities are observed when the carrier density is ~$3 \cdot 10^{17}$ cm$^{-3}$. The mobility decays rapidly as the carrier density increases; this is due to the introduction of extrinsic impurities (Nb, La or oxygen vacancies) associated to potential fluctuations induced in the lattice. On the other hand, for concentrations below $10^{17}$ cm$^{-3}$, the impurity band formed by the donors has a reduced width (due to a decrease of the overlap of the impurity wave functions), and the mobility decreases as well. **Fig. 3** also includes our data for Co-LSTO/STO (Ref. 5.) and LAO/STO sample (this work) which perfectly match the bulk STO data. The idea that the same physical origin governs the behavior of the STO-based heterostructures is further reinforced by the additional recompilation of the data for STO$_{3-\delta}$/STO homoepitaxial films[7] and other LAO/STO samples (Refs. 1-3), considering that carriers extend over a region of 500 μm. *Our third conclusion is therefore that the same physical mechanism is responsible for the conduction and high mobility in all the STO-based systems considered and that this conduction occurs within an extended region of the doped STO substrates.* This conclusion is further confirmed by the observation of the same temperature dependence of the resistance obtained for a sample with the film deposited at high temperature and low oxygen pressure, and that of the same sample after removing *the film and about 5 µm of the STO substrate* by mechanical polishing on the film side.

Different mechanisms have been evoked to explain the high mobilities of LAO/STO samples. Charge redistribution at the LAO/STO interface, preventing the so-called polar catastrophe originated by the polar discontinuity at the LAO/STO interface has been proposed by Nakagawa et al.[10]. If such charge transfer across the interface is the only reason for the reported high-mobility transport properties in LAO/STO interfaces, the latter should not show any thickness nor growth pressure dependence. However, to our knowledge, no highly conductive LAO/STO interfaces have been reported for heterostructures grown at pressures above $10^{-4}$ mbar [1-4]. Furthermore, it was found that the high-mobility behavior actually depends on film thickness and that the carrier densities deduced from Hall effect measurements are much larger than the density of donor sites at the interface[1]. From these considerations and the 3D character of the transport we have emphasized, it seems quite clear that charge transfer plays only a minor role – if any – in the observed high mobility transport properties of LAO/STO samples.

Recently, Siemons et al.[3] suggested that carriers responsible for the high mobility of LAO/STO samples may arise from oxygen vacancies created inside the STO substrate during the growth of the LAO film. They argue that a reduction throughout the bulk of the STO is very unlikely, so that the oxygen vacancies should be confined within a thin layer near the interface to the LAO film. In order to explain the high mobility, they propose that, at low temperature, the carriers stray away from the region of oxygen vacancies, allowing them to have a high mobility. In their picture, the width of the layer in which electrons are distributed has a non-uniform carrier density profile and should depend strongly on temperature. Although this interpretation is compatible with the room-temperature electron energy loss spectroscopy (EELS) data that indicate a gradient of $Ti^{3+}$ extending over 10 - 15 nm from the interface into the STO substrate[11], it is at odds with the experimental observation of well defined SdH frequencies, which hints to a homogeneous electronic system with a uniform carrier density.

We claim that the origin of the high mobility of the LAO/STO systems is the same as in bulk STO, i.e. the homogeneous doping of STO with carriers due to the oxygen vacancies. Our conclusion is also supported by recent cathodo- and photoluminescence studies of Kalabukhov et al. [12]. High mobility values (up to and above $2 \cdot 10^4$ cm$^2$/Vs at low temperatures) for the bulk STO single crystals (uniformly doped throughout their volume with a low amount of Nb-, La- or oxygen vacancies) are known for decades [6, 8, 9, 13, 14, 15, 16, 17]. Such large mobilities are due to the polarization shielding of the ionized scattering centers driven by the large low-temperature dielectric constant of STO in the temperature range where ionized defect scattering is dominant [6].

If the oxygen vacancies are the origin of the high mobility gas, important questions are how they are created in the STO substrate and how deep they can diffuse. A common feature of all the high-mobility STO-based thin film structures is that they are fabricated at low growth pressure and high deposition temperature. On the other hand, to obtain bulk STO single crystals with high mobilities, large exposure times (of the order of days) in reducing atmospheres and high temperature are required [6, 8, 13]. This is also further supported by our findings that a simple annealing of the substrate (with temperature, pressure and time conditions similar to those used during the LAO deposition process) does not yield to a high-conductive material. We conclude, therefore, that the deposition process seems to somehow "catalytically" speed up the surface kinetics so that the exchange of oxygen is strongly enhanced at the film/substrate interface as compared to the substrate/vacuum interface (see, e. g., Ref. 18). Once created, the oxygen vacancies diffuse rather easily inside the STO. Their diffusion coefficient $D_V$ has been measured with different methods[19, 20, 21, 22] and at various oxygen partial pressures and temperatures. For temperatures around 750ºC and at pressures as high as $P_{O2} \approx 10^{-2}$ mbar, values of $10^{-4} - 10^{-5}$ cm$^2$/s were found. During a time interval $t$ the vacancies diffuse along a distance $l_{Ovac} \approx (D_V t)^{1/2}$, which is, for $t = 10$ s (i.e. typical deposition times for the thinnest films) between 100 µm –300 µm. In other words, it is reasonable to assume that oxygen vacancies (created when STO is exposed to reducing atmospheres during the film growth) can extend to a sizeable region within the STO substrate, compatible with that found from the SdH analysis.

In conclusion, we have discussed the characteristics of the high-mobility conducting behavior of LAO/STO samples, and we have given evidences that it is essentially a bulk phenomenon, not different from what has been observed in doped STO bulk samples. We have suggested that these features are the result of the creation and diffusion of oxygen vacancies in the STO substrates during the growth of the LAO film. Our results are strongly supported by the fact that similar transport properties have been observed for a large variety of different systems like doped bulk STO and (if one assumes that carriers extend over the substrate) LAO/STO, CoLSTO/STO and STO$_{3-\delta}$/STO structures, all having in common to be based on STO single crystals exposed to reduced atmosphere at high temperature.

Finally, we want to emphasize the unique potential of heterostructures combining high mobility STO and magnetic oxides for the realization of all-oxide spintronics devices. The LAO layer in the present LAO/STO structures could be replaced by a great variety of magnetic oxides, perovskites or spinels, which can be grown epitaxially on STO, as we have already done with the dilute ferromagnetic oxide CoLSTO [5, 23]. In this way, the spin injection into a high mobility STO channel might be achieved with strongly spin-polarized sources and drains (made of a half-metallic oxide like $La_{2/3}Sr_{1/3}MnO_3$) or by tunnel injection through spin filtering barriers ($BiMnO_3$ [24] or $NiFe_2O_4$ [25]). Additionally, it would be possible to add a large choice of oxides as a gate electrode on a STO channel to either introduce transistor-type functionalities, or to confine spin-polarized electrons into quantum boxes (in the case of thin STO layers). All these possibilities open interesting prospects and should stimulate further studies of all-oxide heterostructures for spintronics.

**METHODS**

We have grown a LAO film of thickness $t = 20$ nm by pulsed laser deposition on a 0.5 mm thick, (001)-oriented and $TiO_2$-terminated STO single-crystalline substrate. The growth was controlled in situ by reflection high-energy electron diffraction. A frequency-tripled ($\lambda=355$ nm) Nd: yttrium aluminum garnet (YAG) laser was used for the growth with a pulse rate of 2.5 Hz and an energy density of 2.8 J/cm$^2$. The growth oxygen pressure was set at $10^{-6}$ mbar and the substrate temperature was 750 °C.

We have characterized the interface by aberration-corrected high-resolution transmission electron microscopy (HRTEM), and found that the LAO layer is fully strained, and the interface is close to atomically sharp, with no dislocations[26].

Transport measurements were performed with an ac setup (22 Hz) and in magnetic fields up to B=16 T and temperatures down to T=1.5 K. For these measurements Al/Au contacts were used on 500 µm x 500 µm pads defined by optical lithography (**Fig. 1e**). The LAO/STO interface was contacted through the LAO layer by locally etching the LAO with accelerated Ar ions down to the interface, in a chamber equipped with a secondary ion mass spectroscopy detection system. We have verified that the Al/Au contacts were ohmic.

Magnetoresistance was measured (at fixed temperatures) for magnetic field oriented either perpendicular (PMR) or parallel (LMR) to the sample plane (and current), and represented as MR = $[R_{xx}(H)-R_{xx}(0)]/R_{xx}(0)$. The Hall resistance $R_{xy}$ was measured at several representative fixed temperatures and in field sweepings from maximum negative to maximum positive value, in order to eliminate the possible mixing of magnetoresistive contribution ($R_{xy}=(R_{xy}(B)-R_{xy}(-B)/2)$. A part of the Hall effect data were obtained from the temperature sweeps of the Hall resistances in 16 T and -16 T. Having confirmed previously the linear field dependence of $R_{xy}(B)$, the Hall mobility at a given temperature was then determined by dividing the Hall resistance at 16 T with the magnetic field and resistance (that was measured independently).

**ACKNOWLEDGMENTS**

G. Herranz acknowledges financial support from the Departament d'Universitats, Recerca i Societat de la Informació de la Generalitat de Catalunya (Spain). The financial support from the PAI-France-Croatia COGITO Program No. 82/240083 is also acknowledged. We acknowledge the experimental support from Y. Lemaître and S. Fusil.

FIGURE 1. **Magnetotransport properties of a LAO/STO bilayer, a,** magnetic field dependence of the perpendicular (blue) and longitudinal (red) magnetoresistances at 1.5 K, **b,** oscillatory part of the magnetoresistance $\Delta R_{SdH}$ obtained by subtracting the background contribution from the measured values, **c,** spectral power density from FFT analysis of $\Delta R_{SdH}(H)$, **d,** the field dependence of the Hall resistance $R_{xy}$ at 4.2 K, **e,** the contact arrangement.

FIGURE 2. **Temperature dependence of the mobility and normalized resistance for different STO systems. a,** Temperature dependence of the Hall mobility $\mu_H$ and **b,** temperature dependence of the resistance normalized to the value at 200 K; both for CoLSTO/STO, LAO/STO, STO$_{3-\delta}$/STO and doped STO single crystals (this work and data from literature). The inset shows the temperature dependence of the resistance ratio $R_{sheet}(300K)/R_{sheet}(2K)$ of the LAO/STO sample reported in this work.

FIGURE 3. **Dependence of the Hall mobility (at 4 K) on the carrier density,** (this work and data from literature): STO single crystals doped with oxygen vacancies (◆) or Nb (◇) [Ref. 8]; slightly reduced STO single crystals (◄) [Ref. 9]; La-doped STO thin films from (►) [Ref. 15]; CoLSTO/STO samples grown at low pressure [Ref. 5 (▲)]; STO$_{3-\delta}$/STO films grown at low pressure (▼) [Ref. 7]; LAO/STO samples grown at low pressure [Ref. 3] (□), [Ref. 1] (■) and this work (●). The dashed line is a guide for the eye. Note that the carrier densities for the data of Refs. 1, 3 and 7, have been recalculated from the original data by assuming that the conductive STO substrate has a thickness of the order of 500 μm.

Fig. 1

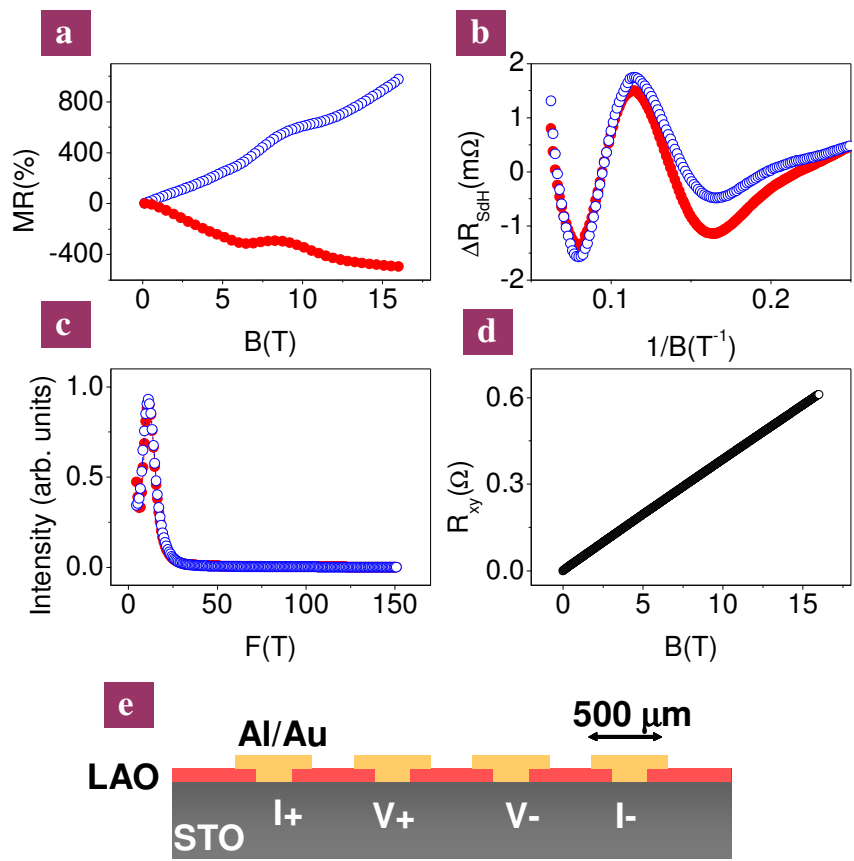

Fig. 2

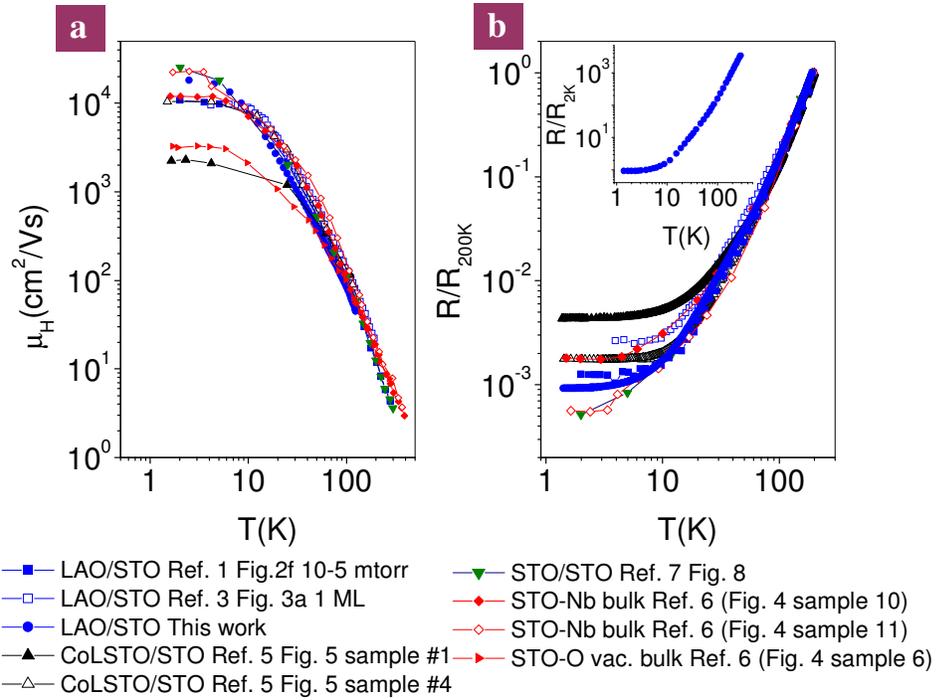

Fig. 3

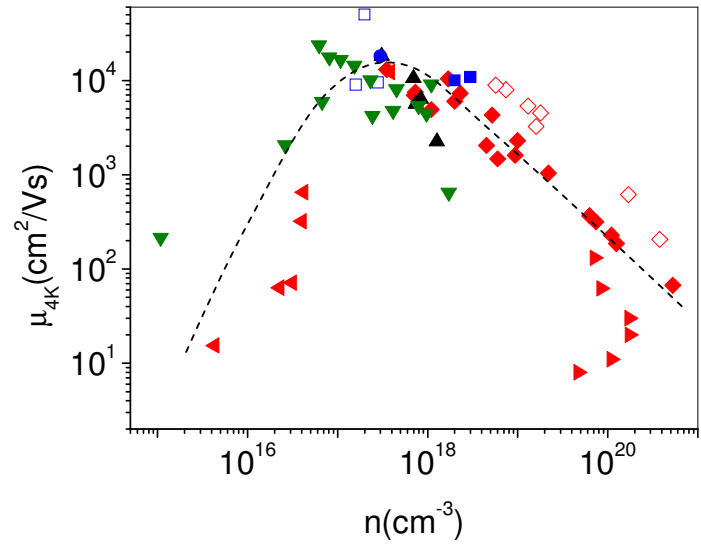

REFERENCES

[1] Ohtomo, A., Hwang, H. Y. A high-mobility electron gas at the LaAlO$_3$/SrTiO$_3$ heterointerface. *Nature* **427**, 423-426 (2004).

[2] Hwang, H.Y., Ohtomo, A., Nakagawa, N., Muller, D.A. & Grazul, J.L. High-mobility electrons in SrTiO$_3$ heterostructures. *Physica E* **22**, 712-716 (2004).

[3] Siemons, W. *et al.* Origin of the unusual transport properties observed at hetero-interfaces of LaAlO$_3$ on SrTiO$_3$. *cond-mat/0603598* (2006).

[4] Huijben, M. *et al.* Electronically coupled complementary interfaces between perovskite band insulators. *cond-mat/0603088* (2006).

[5] Herranz, G. *et al.* Full oxide heterostructure combining a high-T$_C$ diluted ferromagnet with a high-mobility conductor. *Phys. Rev. B* **73**, 064403 (2006).

[6] Tufte, O. N. & Chapman P. W. Electron mobility in semiconducting strontium titanate. *Phys. Rev.* **155**, 796-802 (1967).

[7] Ohtomo, A. & Hwang, H.Y. Growth mode control of the free carrier density in SrTiO$_{3-\delta}$ films. *cond-mat/0604117* (2006).

[8] Frederikse, H. P. R. & Hosler, W. R. Hall mobility in SrTiO$_3$. *Phys. Rev.* **161**, 822-827 (1967).

[9] Lee, C., Yahia, J. & Brebner, J. L. Electronic conduction in slightly reduced strontium titanate at low temperatures. *Phys. Rev. B* **3**, 2525-2533 (1971).

[10] Nakagawa, N., Hwang, H. Y. & Muller, D. A. Why some interfaces cannot be sharp, Nature Mater. *Nature Mater.* **5**, 204-209 (2006).

[11] Maurice, J.-L. Carrétéro, C., Contour, J.-P. & Colliex, C. unpublished.

[12] Kalabukhov, A. S. *et al.* The role of oxygen vacancies in SrTiO$_3$ at the LaAlO$_3$/SrTiO$_3$ interface. *cond-mat/0603501* (2006).

[13] Frederikse, H. P. R., Thurber, W. R. & Hosler, W. R. Electronic transport in strontium titanate *Phys. Rev.* **134**, A442 - A445 (1964).

[14] Frederikse, H. P. R., Hosler, W. R., Thurber, W. R., Babiskin, J. & Siebenmann, P.G. Shubnikov-de Haas effect in SrTiO$_3$. *Phys. Rev.* **158**, 775-778 (1967).

[15] Olaya, D., Pan, F., Rogers, C. T. & Price, J. C. Electrical properties of La-doped strontium titanate thin films. *Appl. Phys. Lett.* **80**, 2928-2930 (2002).

[16] Tomio, T., Miki, H., Tabata, H., Kawai, T. & Kawai, S. Control of electrical conductivity in laser deposited SrTiO$_3$ thin films with Nb doping. *J. Appl. Phys.* **76**, 5886-5890 (1994).

[17] Leitner, A., Rogers, C. T., Price, J. C., Rudman, D. A. & Herman, D. R. Pulsed laser deposition of superconducting Nb-doped strontium titanate thin films. *Appl. Phys. Lett.* **72**, 3065-3067 (1998).

[18] Leonhardt, M., De Souza, R.A., Claus, J. & Maier, J., Surface kinetics of oxygen incorporation into SrTiO$_3$, *J. Electrochem Soc.* **149**, J19-J26 (2002).

[19] Schwarz, D.B. & Anderson, H.U. Determination of oxygen chemical diffusion coefficients in single crystal SrTiO$_3$ by capacitance manometry *J.Electrochem.Soc* **122**, 707-710 (1975).

[20] Paladino, A.E. Oxidation kinetics of single-crystal SrTiO$_3$. *J.Am.Ceram.Soc*. **48**, 476-478 (1965).

[21] Ishigaki, T., Yamauchi, S., Kishio, K., Mizusaki, J. & Fueki, K. Diffusion of oxide ion vacancies in perovskite-type oxides. *J.Solid State Chem.* **73**, 179-187 (1988).

[22] Denk, I., Noll, F. & Maier, J. In situ profiles of oxygen diffusion in SrTiO$_3$: bulk behavior and boundary effects. *J. Am. Ceram. Soc.* **80**, 279 (1997).

[23] Herranz, G. *et al.* Co-doped (La,Sr)TiO$_{3-\delta}$: A High Curie Temperature Diluted Magnetic System with Large Spin Polarization. *Phys. Rev. Lett.* **96**, 027207 (2006).

[24] Gajek, M. *et al.* Spin filtering through ferromagnetic BiMnO$_3$ tunnel barriers. *Phys. Rev. B* **72**, 020406 (2005)

[25] Lüders, U. *et al.* Spin filtering through ferrimagnetic NiFe$_2$O$_4$ tunnel barriers. Appl. Phys. Lett. **88**, 082505 (2006)

[26] Maurice, J.-L. *et al.*, Electronic conductivity and structural distortion at the interface between insulators SrTiO$_3$ and LaAlO$_3$. *Phys. Stat. Sol. (a)* in press, available at *cond-mat/0511123* (2006).